\documentclass[a4paper]{article}

\usepackage{amstex}
\usepackage{amssymb}
\usepackage{apalike}

\makeatletter
\newcommand{\mysection}%
{\@startsection{mysection}{1}{0em}{-\baselineskip}%
{\baselineskip}{\large\bfseries}}
\makeatother

\newcommand{\preprintdate}{April 1996}
\newcommand{\preprintnumber}{IFUP--TH 22/96}
\newcommand{\hepnumber}{hep-ph/9604361}
\newcommand{\titletext}{Electric dipole moments in supersymmetric theories}
\newcommand{\authortext}{
\textbf{Andrea Romanino}%
\thanks{Email: romanino\symbol{64}ibmth.difi.unipi.it}
\\\em Dipartimento di Fisica, Universit\`a di Pisa and
INFN, Sezione di Pisa, I-56126 Pisa, Italy}
\newcommand{\abstracttext}{%
Intrinsic EDMs in microscopic systems at a level of sensitivity 
achievable in
experiments under way or foreseen are predicted in supersymmetric unified
theories. I describe this and other sources of measurable EDMs and I
show how these sources can be distinguished through experiments in
different systems.}

\newcommand{\makeonlyfirstpage}{%
\setlength{\topmargin}{21mm}
\setlength{\textheight}{172mm}
\renewcommand{\footnoterule}{} 
\title{%
\normalsize\hspace*{\fill}
\begin{tabular}{l}\preprintnumber\\\hepnumber\end{tabular}
\vspace{3\baselineskip}\\\huge\bfseries\titletext}
\author{
\begin{minipage}[t]{0.8\textwidth}
\Large\centering\authortext
\end{minipage}}
\date{\preprintdate}
\begin{document}
\maketitle
\begin{abstract}\large\noindent\abstracttext\end{abstract}
\thispagestyle{empty}
\end{document}}

\newcommand{\TeV}{\,\mathrm{TeV}}
\newcommand{\ecm}{e\,\mathrm{cm}}
\newcommand{\mutoegamma}{\mu\rightarrow e\gamma}
\newcommand{\M}[1]{M_{\mathrm{#1}}}
\newcommand{\fig}[1]{Fig.\ref{#1}}

\title{\huge\bfseries\titletext}
\author{\begin{minipage}[t]{0.8\textwidth}
\large\centering\authortext
\end{minipage}}
\date{}

\begin{document}

\maketitle
\begin{abstract}\large\noindent\abstracttext\end{abstract}
\vspace{\baselineskip}

\noindent
The intrinsic dipole moments (EDMs) of microscopic systems constitute
an observable of great interest because they can provide
informations on CP violation and physics beyond the Standard Model.
An improvement of the sensitivity of the present experiments by one or
two orders of magnitude, as pursued by foreseen experiments, might
give rise to a positive signal \cite{ramsey:95a,demille:95a}. 
The present limits for the neutron \cite{altarev:92a} and
electron\footnote{The EDM is measured in neutral systems; on the
other side, the EDM of paramagnetic atoms is mainly due to the EDM of
the unpaired electron \cite[and references
therein]{barr:92a,fischler:92a} and this allows for an indirect
determination of the electron EDM.} \cite{commins:94a} EDM are shown
below
\begin{equation}
d_n<0.8\cdot 10^{-25}\ecm \qquad d_e<4.3\cdot 10^{-27}\ecm.
\end{equation}

EDMs in this range are predicted as unavoidable effects\footnote{Barring
accidental cancellations.} in presence of supersymmetric
unification with SO(10) as gauge group if the supersymmetry breaking
is communicated to the observable fields at the Planck scale. 
From this point of view,
supersymmetric unification differs from non unified and/or non
supersymmetric models in which the CP violation is due to the CKM
phase, that all predict much too small
EDMs.
A possible EDM signal in experiments in progress or foreseen could then
represent, as $\mutoegamma$ and other flavour violating processes
\cite{barbieri:94a,barbieri:95a}, an indirect signal of supersymmetric
unification \cite{dimopoulos:95a,barbieri:95b}.
However, the same effect can also be generated in non unified
theories, in particular in the Minimal Supersymmetric Standard Model
(MSSM) if the soft supersymmetry breaking parameters are complex
already at the Planck scale, or, more generally, if
strong CP violation is present.

In this presentation, I mean first to discuss the origin of the
large effect in unified supersymmetric models and then to consider the
possibility of distinguishing experimentally among the quoted sources of
CP violation at the level of present or foreseen experiments (in
the following, I will use ``observable effects'' with this meaning).

\mysection*{EDMs in superunified models}

EDMs, as every signal of CP violation, require the presence of
uneliminable phases in the lagrangian of the theory.
In the Standard Model, the only phase is contained in the CKM matrix
that appears in the charged current interactions expressed in terms of
mass eigenstates. The effects of this CKM phase are largely
unobservable: the prediction for the neutron EDM is\enspace
$d_n\lesssim\mathcal{O}(10^{-33})\ecm$ \cite{shabalin:83a}.

In a generic supersymmetric extension of the Standard Model, the
presence of new fields and couplings  can
involve the existence of new uneliminable phases. In particular they
can arise when the couplings of quarks and leptons with their
supersymmetric partners are expressed in terms of mass eigenstates and
the corresponding transition matrices appear in the couplings.

Let us suppose that the soft supersymmetry breaking parameters are
flavour universal
and real at the Planck scale $\M{P}$, as it happens in a large class
of models in which the supersymmetry breaking is communicated to the
observable sector by gravity \cite{barbieri:82a,chamseddine:82a}.
In this case, at the Planck scale and at the tree level the scalar
masses are degenerate in flavour, so no transition matrices appear in
the fermion-sfermion couplings. However, the non universal couplings of
the theory, namely the Yukawa matrices, correct radiatively the soft
parameters, that lose in this way their universality. In particular,
at the Fermi
scale, where the EDMs are generated, the corrections due to the top
Yukawa coupling, that we know to be large, are important. If the ratio
of the vacuum expectation values of down and up Higgs is small (large
$\tan\beta$), also the bottom Yukawa coupling can be important.

Is this enough in order to have observable EDMs?
The answer depends on the theory structure below the Planck scale.
More precisely, if the gauge group is the Standard Model one until the
Planck scale, the effects are well below the experimentally accessible
region, whereas, if an
intermediate stage of unification is present between the Fermi scale
and the Planck scale, observable effects can be generated both for the
neutron and for the electron.

The reason for this difference is that in the former case the EDMs
due to the radiatively induced fermion-sfermion
mixing are vanishing when some of the light Yukawa couplings  are
vanishing, whereas this
does not happen in the latter case.
In fact, let us consider for example the EDM of the down quark in the
limit of vanishing up and charm Yukawa couplings $\lambda_u =
\lambda_c = 0$.
If the Standard Model gauge group holds until the Planck scale, in
this limit there is no CP violation in the theory and the $d$ quark EDM
is vanishing.
In fact, at the Planck scale the soft parameters are universal and
real by hypothesis and, after eliminating non physical phases from the
CKM matrix, the
physical one can be reached and eliminated with a U(2) rotation in the
up-charm sector and a redefinition of field phases. On the other hand,
if at the Planck scale the theory is unified, it is impossible to do a
rotation or a phase redefinition only in the left sector, because
this sector is unified with the up right one.

Let us look now more closely how EDMs for the $d$ quark and the electron
arise in the case of SO(10) unification and moderate $\tan\beta$, in
which the only important Yukawa is the top one $\lambda_t$.
At the Planck scale and at the tree level, in the universality
hypothesis the supersymmetry breaking mass term for the scalars of the
three generations is 
\(
m^2_{\mathrm{P}}\tilde{16}_i^{\dagger}\tilde{16}_i
\),
where $\tilde{16}_i$ is the field that unifies the scalar-quarks and
scalar-leptons of the $i$\emph{-th} generation. As a consequence of radiative
corrections due to $\lambda_t$, at the unification scale $\M{G}$ the
mass of \emph{all} third generation scalars is different from the mass
of other generations and, in the basis in which the Yukawa matrix that
involve $\lambda_t$ is diagonal, the mass term becomes
\begin{equation}
\sum_{i=1}^2
m^2_{\mathrm{G}}\tilde{16}_i^{\dagger}\tilde{16}_i
+(m^2_{\mathrm{G}} -\delta
m^2_{\mathrm{G}})\tilde{16}_3^{\dagger}\tilde{16}_3.
\end{equation}
From the GUT scale to the Fermi scale, the scalar partners of the top,
left and right, and of the bottom left are further corrected.

In the basis considered both the up and scalar up mass
matrices are diagonal, whereas the down and charged lepton mass matrices
are non diagonal. Therefore, the relative rotation between the mass
eigenstates of fermions and scalars is $\mathbf{1}$ in the up sector
(both left and right), whereas in the down and charged lepton sectors
it is given by the CKM matrix\footnote{More precisely, except for
the down left sector, the rotation is given by the CKM matrix
evaluated at the unification scale.}.
The misalignment in the up sector and the resulting up quark EDM can
be important in the large $\tan\beta$ case. In the present case only
down quark and electron EDMs are generated in a significant way.
In a wide region of parameters space, EDMs in the observable range are
predicted for the neutron and the electron
\cite{dimopoulos:95a,barbieri:95a,barbieri:95b}.

The leading contributions to the EDMs of $d$ and $e$ turn out to be
proportional to the bottom
and tau mass respectively through the insertion of the left-right
block of the scalar mass matrices.
They are also proportional to the imaginary part of a product of CKM
matrix elements that involves $V_{td}^2$.
Moreover, they grow as $\lambda_t^4$ because the effect is a
consequence of the splitting in the scalar mass matrices both in the
left and right sector.
More precisely, $d_d$ is
subject to a GIM suppression if $\delta m^2_{\tilde{d}_L}/M_3^2$ or
$\delta m^2_{\tilde{d}_R}/M_3^2$ is small, where $M_3$ is the gluino
mass and $\delta m^2_{\tilde{d}_L}$, $\delta m^2_{\tilde{d}_R}$ are
the 1-3 splittings in the mass of left and right down scalars, both
proportional to $\lambda_t^2$. The
contribution to $d_e$ is subject to the same kind of suppression if
$\delta m^2_{\tilde{e}_L}/M_2^2$ or $\delta m^2_{\tilde{e}_R}/M_2^2$
is small, where this time $M_2$ is the $\text{SU(2)}_L$ gaugino mass.
Since at the Fermi scale $M_3^2$ is significantly greater than $M_2^2$,
whereas the 1-3 splittings in the squark and slepton sectors are not so
different, the GIM cancellation is
more effective in suppressing the down EDM relative to the electron
one.

\mysection*{Comparison among different sources}

We saw that supersymmetric unification is a source of EDMs observable
in experiments under way or foreseen.
However, a signal, e.g., of a neutron EDM does not constitute
``per se'' an evidence of supersymmetric unification.
As a matter of fact, also in the MSSM with universal soft terms,
observable effects can arise if the soft supersymmetry breaking
parameters are complex already
at the Planck scale
\cite{polchinski:83a,aguila:83a,buchmuller:83a}. In this case, we can
have two further
physical phases that, in a supersymmetric basis, are those of the B
term and of the universal A term at the Planck scale (defined as the
parameters that multiply the corresponding superpotential parameters
in the bilinear and trilinear scalar interactions respectively).
These phases give rise to large EDMs for the electron, the down quark
and also the up quark, because they appear both in the left-right
block of scalar mass matrices and in the gaugino-higgsino mass
matrices.
In this case the EDMs of $e$, $d$ and $u$ are proportional to their
masses unlike the previous case where the fermion-sfermion mixing
allowed to have third generations in the scalar loops.
Moreover, another possibility that has to be considered is that the possible
signal of an EDM is a strong CP violation effect.

In the event that EDMs will be measured in the near future, is
therefore important to be able to distinguish among the described
sources, the superunification, the presence of soft universal phases
and
strong CP violation. It turns out that the range of values
that the neutron and electron EDMs $d_n$, $d_e$ can individually
assume in the three cases are largely overlapping themselves, due to
the unknown phases. On the other hand, the
predictions for the ratio $d_n/d_e$ are characteristic
of the different source \cite{barbieri:96a}.

Quite clearly, if the effect were due to strong CP violation, only the
neutron would have a significant EDM. Moreover, in the case of EDMs
generated by soft phases the $d_n/d_e$ ratio is greater than in the
unification case for two reasons: first, as we saw, in the latter
case there is a GIM cancellation that is more effective in suppressing
$d_d$ and hence $d_n$\footnote{In addition, in this case the up quark
does not contribute to $d_n$, but, from a numerical point of view,
this is not very important.} relative to $d_e$. Furthermore, in the
former case the fermion EDMs are proportional to their masses, whereas
in the latter case they are proportional to the corresponding third
generation fermions, so that
\begin{equation}
\fracwithdelims(){d_n}{d_e}
\begin{Sb} \mathrm{soft~~~}\\\mathrm{phases} \end{Sb}
\text{\LARGE /}
\fracwithdelims(){d_n}{d_e}_{\mathrm{GUT}}
\propto
\frac{m_d}{m_e}\text{\large /}\frac{m_b}{m_{\tau}}\approx 5 \pm 2.
\end{equation}

These qualitative considerations have been made quantitative in a
numerical analysis in which the contributions of chromoelectric dipole
moments of quarks to the neutron EDM are taken in account (the
Weinberg three-gluon operator and four-fermion operators do not
contribute in a significant way) and the parameters are made to vary
in a random way \cite{barbieri:96a}. From this
analysis it is confirmed
that $d_n$ or $d_e$ generated by soft phases are generally not
consistent with the
present experimental limits in most of relevant parameter space unless
the phases are small, whereas this is not the case in superunification.
Irrespective of the values of the phases, the $d_n/d_e$
ratio is actually smaller in the unification case: $0.12\cdot 10^{\pm
0.83}$ against $4.2\cdot 10^{\pm 0.34}$ in the other case\footnote{The
errors come from the random sampling in the parameter space.}. The two
predictions are closer when  the gluino is light, because in this
region the GIM suppression of $d_d$ fails.\vspace{\baselineskip}

To conclude, current and/or foreseen experiments could measure an EDM
for the neutron and/or electron and therefore provide evidence for CP
violation and physics beyond the Standard Model. Such signals must be
considered likely if supersymmetric unification is realized in
nature. Furthermore, the
combined effort of experiments on various systems can provide crucial
information to distinguish the phisical source of the signal(s).

\end{document}